\documentclass[conference,english,a4paper]{IEEEtran}
\usepackage[T1]{fontenc}
\usepackage[latin1]{inputenc}
\usepackage{graphicx}
\usepackage{amssymb}
\usepackage{amsmath}

\makeatletter

\providecommand{\boldsymbol}[1]{\mbox{\boldmath $#1$}}

\newtheorem{thm}{Theorem}

\usepackage{babel}
\makeatother

\begin{document}

\title{Distributed Space-Time Block Codes for the MIMO Multiple Access Channel}

\author{\authorblockN{Maya Badr}\authorblockA{E.N.S.T.\\46, rue Barrault\\75013 Paris, France\\Email: \texttt{mbadr@enst.fr}}\and\authorblockN{Jean-Claude Belfiore}\authorblockA{E.N.S.T.\\46, rue Barrault\\75013 Paris, France\\Email: \texttt{belfiore@enst.fr}}}

\maketitle
\begin{abstract}
In this work, the Multiple transmit antennas Multiple Access Channel is considered. A construction of a family of distributed space-time codes for this channel is proposed. No Channel Side Information at the transmitters is assumed and users are not allowed to cooperate together. It is shown that the proposed code achieves the Diversity Multiplexing Tradeoff of the channel. As an example, we consider the two-user MIMO-MAC channel. Simulation results show the significant gain offered by the new coding scheme compared to an orthogonal transmission scheme, \textit{e.g.} time sharing. 
\end{abstract}

\section{Introduction}
Multiantenna Multiple Access Channel (MIMO-MAC) has recently received a great interest but optimal and practical design of space-time codes for this channel is still missing. After introducing the diversity-multiplexing tradeoff (DMT) of a MIMO channel in \cite{Zheng-1}, Tse \emph{et al.} introduced the DMT of the MAC in \cite{DMTMAC}. It is a fundamental limit of the channel at high $\mathsf{SNR}$ that can be used to evaluate the performance of different transmission schemes. 
Still in \cite{DMTMAC}, they proved that this DMT is achievable for sufficiently long codes by considering a family of Gaussian random codes. However, these Gaussian codes don't have any structure which makes their efficient encoding and decoding impractical. 

Nam and El Gamal proposed in \cite{elgamal-mac}, a class of structured multiple access lattice space-time codes. They proved that their scheme, based on lattice decoding, achieves the optimal DMT of the MAC but they did not give any constructive example. This fact lets us think that, as it was the case for the MIMO channel \cite{ElGamal-4}, the construction of such lattice codes for the MAC should use large alphabets with prime cardinality.

In \cite{helmut}, G\"artner and B\"olcskei presented a detailed analysis of the MAC based on the different error types that can be encountered in this channel (see \cite{gallager}). They derived a space-time code design criterion for multiantenna MACs and presented a structured coding scheme of length $4$ for $2$ transmit users with two transmit and two receive antennas. This code results from a simple concatenation of two Alamouti codewords with a columns swapping for one user's codeword offering a minimum rank of three. This code highlights the importance of the joint code design in the MAC, but does not achieve the outage DMT of the channel.
G\"artner and B\"olcskei further developped their work and presented in \cite{Helmut-new} important results motivating the construction of space-time codes for the MIMO-MAC. They showed that their code design criteria are optimal with respect to the DMT of the channel and proved that, for a MIMO-MAC, outage analysis allows a rigorous characterization of the dominant error event regions. In other words, outage and error probabilities have the same behaviors at high $\mathsf{SNR}$. This fundamental result will be of a major importance in our work.

In this paper, we present a construction of a family of distributed space-time codes for the MIMO-MAC with no Channel Side Information at the transmitters (no CSIT) based on the fundamental results in \cite{Helmut-new}. By analyzing the different regimes of the DMT of the MIMO-MAC, we show that the new codes achieve the outage DMT of the channel. Numerical results finally show that the proposed codes outperform the time sharing scheme and that the error probabilities of such codes mimic the outage probability behavior of the MIMO-MAC channel. In the sequel, we first present a general codes construction, \textit{i.e.} for the $(K, n_t, n_r)$ MIMO-MAC. Then, as a detailed example, we consider the two-user MAC case with $n_t=2$.

\section{The Multi-antenna Multiple-Access Channel}
\subsection{System model}
In this paper, we use boldface capital letters $\boldsymbol{M}$ to denote matrices. $\mathcal{CN}$ represents the complex Gaussian random variable. [.]$^\top$ (\textit{resp.} [.]$^\dagger$) denotes the matrix transposition (\textit{resp.} conjugated transposition) operation. 

We consider a $K$-user multiple-access channel with $n_{t}$ transmit antennas per user and $n_{r}$ receive antennas. We assume that the channel matrices have \textit{i.i.d.} zero-mean Gaussian entries, \textit{i.e.}, $h_{i,j} \sim \mathcal{CN} (0,1)$. We denote $T$ the temporal codelength of the considered distributed space-time code $\mathcal{C}$. Let $\boldsymbol{X}_{i}$ be a $\left(n_{t}\times T\right)$ matrix denoting the codeword of user $i$ with normalized power, independent of codewords of the other users since we assume no cooperation between users. The received signal is \begin{equation}
\boldsymbol{Y}^{\left(n_{r}\times T\right)}=\sum_{i=1}^{K}\boldsymbol{H}_{i}^{\left(n_{r}\times n_{t}\right)}\boldsymbol{X}_{i}^{\left(n_{t}\times T\right)}+\boldsymbol{W}^{\left(n_{r}\times T\right)}\label{eq:rec-sig}\end{equation}
where superscripts denote matrices dimensions. $\boldsymbol{W}$ is the additive white Gaussian noise matrix with \textit{i.i.d.} Gaussian unit variance entries, \textit{i.e.}, $\boldsymbol{W} \sim \mathcal{CN}(0,1)$.

\subsection{Diversity-Multiplexing tradeoff interpretation}\label{seq:dmt}
\begin{figure}
\centering
\includegraphics[width=0.8\linewidth]{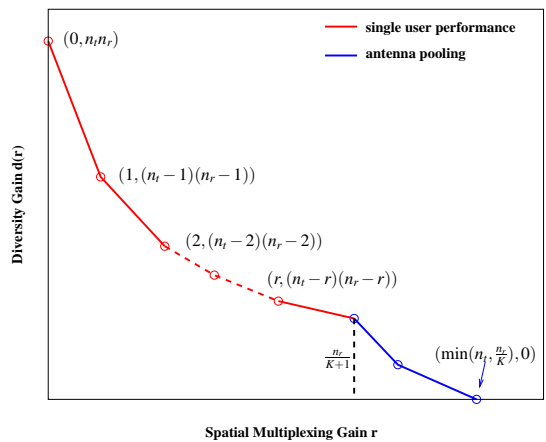}
\caption{DMT of a multiple-access channel with $K$ users with $n_t$ transmit antennas each and a single receiver with $n_r$ antennas.}
\label{fig:DMTgen}
\end{figure}

The diversity-multiplexing tradeoff (DMT) of multiple access channels was introduced and fully characterized in \cite{DMTMAC}. A scheme $\mathcal{C}(\mathsf{SNR})$ is said to achieve multiplexing gain $r$ and diversity gain $d$ if\begin{eqnarray*}
\lim_{\mathsf{SNR}\rightarrow\infty}\frac{R(\mathsf{SNR})}{\log\mathsf{SNR}}=r & \mbox{and} & \lim_{\mathsf{SNR}\rightarrow\infty}\frac{\log P_{e}(\mathsf{SNR})}{\log\mathsf{SNR}}=-d
\end{eqnarray*}

where $R(\mathsf{SNR})$ and $P_{e}(\mathsf{SNR})$ denote respectively the data rate as a function of $\mathsf{SNR}$, measured in bits per channel use (BPCU), and the block error probability. Tse \emph{et al.} gave in \cite{DMTMAC} the optimal achievable tradeoff $d^{\star}(r)$ of the MAC channel which corresponds to the $\mathsf{SNR}$ exponent of the outage probability. The network is assumed to be \textit{symmetric}, \textit{i.e.}, the diversity orders and the multiplexing gains per user are identical ($r$ and $d$). The authors distinguished two loading regimes: the lightly loaded regime, \textit{i.e.} $r \leq \min(n_t,\frac{n_r}{K+1})$, and the heavily loaded regime, \textit{i.e.} $r \geq \min(n_t,\frac{n_r}{K+1})$. In the first regime, single-user performance is achieved, in other words, the presence of other users does not influence the channel performance, whereas in the second one, the system is equivalent to a MIMO system as if the $K$ users pooled up their transmit antennas together. The global DMT is shown to be the minimum DMT between these two regimes, that is, the largest achievable symmetric diversity gain for fixed symmetric multiplexing gain,
\begin{equation}\label{eq:dmt_min}
d^{*}_{sym}(r)=\min_{k=1,\dots,K} d^{*}_{kn_t,n_r}(kr)
\end{equation}
We have the following result\footnote{In the sequel, $d^{*}_{n_t,n_r}(r)$ denotes the outage DMT of a $n_t \times n_r$ MIMO Rayleigh point-to-point channel.} illustrated in figure \ref{fig:DMTgen}
\begin{equation}\label{eq:dmt_diffr}
\centering
d^{*}_{sym}(r) =
\left\{ \begin{array}{ll}
d^{*}_{n_t,n_r}(r), ~~~~~~ r \leq \min(n_t,\frac{n_r}{K+1})\\
d^{*}_{Kn_t,n_r}(Kr), ~~~ r \geq \min(n_t,\frac{n_r}{K+1})
\end{array} \right.
\end{equation}

It is noteworthy, though, that while for $n_r \geq (K+1)n_t$ the use of a code designed for the single-user MIMO channel is optimal, a jointly designed code will be of major importance in the case of $n_r \leq (K+1)n_t$. In fact, depending on the number of receive antennas, the antenna pooling regime may or may not exist: if $n_r \geq (K+1)n_t$, single-user performance is achieved for all $r$ and optimal space-time codes designed for the single-user MIMO channel achieve the outage DMT, else, both the single user regime and the antenna pooling regime occur and should be taken into account in the code design.

\subsection{Code design criteria}
G\"artner and B\"olcskei \cite{helmut,Helmut-new} used an error event analysis, which was first introduced by Gallager in \cite{gallager}, to establish the space-time code design criteria for the MAC. Such an approach consists in defining different error events, say event $\boxed{i}$, depending on the number of users in error\footnote{Event $\boxed{i}$ means $i$ users in error}. Different error regions are defined based on the users' transmission rate. Rate regions where single-user error events dominate can be treated by using well known space-time codes designed for the single-user case. However, the rate regions where the event of more than one user being in error dominates, require a joint code design.

It is interesting to recall the following result presented in \cite{helmut}: increasing the number of receive antennas results in a reduction of the size of the region where all users are in error and thus, decreases the importance of the joint code design. Interestingly, this result confirms the previous DMT interpretation. 

The code design criteria that we use are derived in \cite{helmut}, based on the dominant error regions, using a refined upper bounded expression of the pairwise error probability (PEP) and can be stated as follows
\noindent
\begin{enumerate}
\item \textit{Rank criterion}: For every codeword pair $(\boldsymbol{X}_k,\boldsymbol{Y}_k)$ with $\boldsymbol{X}_k \neq \boldsymbol{Y}_k$ the rank of the corresponding codeword difference matrix shall be maximized.
\item \textit{Eigenvalue criterion}: For every codeword pair $(\boldsymbol{X}_k,\boldsymbol{Y}_k)$ with $\boldsymbol{X}_k \neq \boldsymbol{Y}_k$ the product of the nonzero eigenvalues of the corresponding codeword difference matrix shall be maximized.
\end{enumerate}

Authors further showed that their space-time code design criteria are optimal with respect to the entire DMT and concluded that, a rigorous characterization of the dominant error event regions can be obtained by analysing the outage-DMT of the MAC.

\section{Construction of distributed space-time codes for the MIMO MAC}

The general $(K, n_t, n_r)$ MIMO MAC is considered in this section. We construct a new family of space-time codes for this channel following the same footsteps as the construction of perfect space-time codes for parallel MIMO channels in \cite{sheng}. We assume that the modulation used by both users is a quadrature amplitude modulation (QAM).\footnote{Generalization to hexagonal (HEX) modulation is straightforward.}

\subsection*{Code construction}
Let $\mathbb{F}$ be a Galois extension of degree $K$ on $\mathbb{Q}(i)$ with Galois group
$$\mathrm{Gal}(\mathbb{F}/\mathbb{Q}(i))=\{\tau_1, \tau_2, \dots, \tau_K\}.$$ We denote $\mathbb{K}$ a cyclic extension of degree $n_t$ on $\mathbb{F}$ and $\sigma$ the generator of its Galois group, $\mathrm{Gal}(\mathbb{K}/\mathbb{F})$. Let $\eta$ be in $\mathbb{F}$ such that $\eta,\eta^2,\dots,\eta^{n_t-1}$ are not norms in $\mathbb{K}$. A cyclic division algebra of degree $n_t$ is constructed, $\mathcal{A}=(\mathbb{K}/\mathbb{F}, \sigma, \eta)$. To remind the most relevant concepts about cyclic algebras and how to use them to build space-time block codes, we let the reader refer to \cite{booklet}. We denote $\boldsymbol{\Xi}$ the matrix representation of elements of $\mathcal{A}$ which is a $n_t \times n_t$ matrix and we construct codewords as follows \begin{equation}
\boldsymbol{X}_k=\left[\begin{array}{cccc}
\tau_1(\boldsymbol{\Xi}_k) & \tau_2(\boldsymbol{\Xi}_k) & \dots & \tau_K(\boldsymbol{\Xi}_k)\end{array}\right]\label{eq:codeword_gen_per_user}
\end{equation}

Each user sends its information by transmitting a matrix of the same type as in (\ref{eq:codeword_gen_per_user}), say $\boldsymbol{X}_{k}$ for user $k$. The equivalent joint codeword matrix can be written as \begin{equation}
\boldsymbol{X}=\left[\begin{array}{cc}
\boldsymbol{X}_{1} \\
\boldsymbol{X}_{2} \\
\vdots \\
\boldsymbol{X}_{K} \end{array}\right]\label{eq:codeword-eq-gen}
\end{equation}

Such a code uses $K.n_{t}^{2}$ information symbols per user. In order to check the rank design criterion given in \cite{helmut}, we need to insert in eq. (\ref{eq:codeword_gen_per_user}) a carefully chosen matrix $\boldsymbol{\Gamma}$ so that the transmitted codeword of (\ref{eq:codeword-eq-gen}) be of full rank. We propose the new code $\mathcal{C}_{(K,n_t,\boldsymbol{\Gamma})}$ where each user codeword is
\begin{equation}
\boldsymbol{X}_k=\left[\begin{array}{cccc}
\boldsymbol{\Gamma}\tau_1(\boldsymbol{\Xi}_k) & \boldsymbol{\Gamma}\tau_2(\boldsymbol{\Xi}_k) & \dots & \tau_K(\boldsymbol{\Xi}_k)
\end{array}\right]\label{eq:codeword_gen_per_user_sigma}
\end{equation}
where $\boldsymbol{\Gamma}$ is a multiplication matrix factor for the $k-1$ first matrices of $\boldsymbol{X}_k$. 

We choose $\Gamma\in \mathcal{A}$ with entries in $\mathbb{Q}(i)$ and such that $\det(\boldsymbol{X}) \neq 0$ for all $\boldsymbol{\Xi}_k \neq 0$. With such a code, we can state,
\begin{thm}
$\mathcal{C}_{(K,n_t,\boldsymbol{\Gamma})}$ achieves the outage DMT of the MIMO-MAC 
\end{thm}
\begin{proof} We give here a sketch of proof, details are omitted for lenght constraint. The idea is to prove that by scaling the size of the underlying QAM constellations by a factor of $\mathsf{SNR}^{r}$, the exponent of $\mathsf{SNR}$ in the asymptotic expression of the error varies as the optimal DMT.

If only one user, say $k$, is in error the receiver can cancel signals it receives from the other $K-1$ users and the system is equivalent to a single-user $n_t \times n_r$ MIMO system. In this case, the transmitted codeword $\boldsymbol{X}_k$ is given in (\ref{eq:codeword_gen_per_user_sigma}). The code is equivalent to well-known codes constructed on cyclic division algebras and thus is DMT achieving, \cite{elia}.

If all users are in error, the system is equivalent to an $Kn_t \times n_r$ MIMO channel and the transmitted codewords are given in (\ref{eq:codeword-eq-gen}). In order to preserve the shaping of the code, matrix $\boldsymbol{\Gamma}$ should be unitary. The DMT achievability is guaranteed by the carefull choice of $\boldsymbol{\Gamma}$. For example, we can choose $\boldsymbol{\Gamma}=\gamma \boldsymbol{I}_{n_t}$ where $\gamma$ is a transcendantal number. In that case, the determinant of a codeword, which is a polynomial function of $\gamma$ with coefficients in $\mathbb{K}$, is non zero. As it is proven in \cite{Helmut-new}, this result is sufficient to prove the outage DMT achievability. 
\end{proof}

\section{An example: $K=2, n_t=2$}

\begin{figure*}[!t]
\normalsize
\begin{equation}
\boldsymbol{\Xi}_k=\frac{1}{\sqrt{5}}\left[\begin{array}{cc}
\alpha . (s_{k,1}+s_{k,2}\zeta_8+s_{k,3}\theta+s_{k,4}\zeta_8\theta) & \alpha . (s_{k,5}+s_{k,6}\zeta_8+s_{k,7}\theta+s_{k,8}\zeta_8\theta)\\
\zeta_8 \bar{\alpha} . (s_{k,5}+s_{k,6}\zeta_8+s_{k,7}\bar{\theta}+s_{k,8}\zeta_8\bar{\theta}) & \bar{\alpha} . (s_{k,1}+s_{k,2}\zeta_8+s_{k,3}\bar{\theta}+s_{k,4}\zeta_8\bar{\theta})\end{array}\right]\label{eq:xi}
\end{equation}
\hrulefill
\vspace*{4pt}
\end{figure*}

\subsection{Optimal DMT}\label{sec:ex-opt-DMT}
\begin{figure}[ht]
\noindent \begin{centering}
\includegraphics[width=0.85\columnwidth,keepaspectratio]{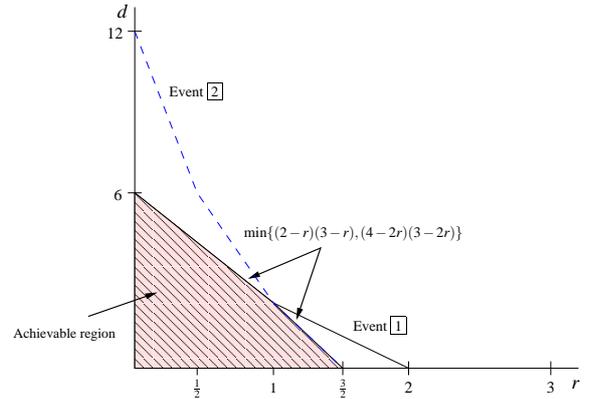}
\par\end{centering}
\caption{\label{fig:DMT-223}DMT of the $(K=2,n_{t}=2,n_{r}=3)$ MAC}
\end{figure}

As an example, we consider a two-user MAC with two transmit antennas per user and three receive antennas, \textit{i.e.}, $n_t=2$ and $n_r=3$.
Based on (\ref{eq:dmt_min}), we can write the optimal DMT in this scenario as follows
\begin{equation}
d^{\star}(r) =  \min\left\{ d^{*}_{2,3}(r), d^{*}_{4,3}(2r)\right\}
\end{equation}
This outage-DMT is illustrated in figure \ref{fig:DMT-223}. Two outage events, leading to the achievable region, are observed: event $\boxed{1}$ where only one user is in outage, the other one being perfectly decoded at the receiver and event $\boxed{2}$ when both users are in outage.

Our goal is to design distributed space-time codes for this channel that are optimal in the sense of the DMT. In other words, the error probability of the proposed scheme should behave asymptotically as the outage probability of the channel. 

\subsection{Code construction}
For the two-user MAC with $n_t=2$, we propose the following code. Each user's codeword is an $2 \times 2$ matrix constructed as follows. Let $\mathbb{F}=\mathbb{Q}(\zeta_8)$ be an extension of $\mathbb{Q}(i)$ of degree $K=2$, with $\zeta_8=e^{\frac{i \pi}{4}}$ and $\mathbb{K}=\mathbb{F}(\sqrt{5})=\mathbb{Q}(\zeta_8, \sqrt{5})$. As explained in \cite{sheng}, such a design leads to construct the Golden code \cite{GC} on the base field $\mathbb{Q}(\zeta_8)$ instead of the base field $\mathbb{Q}(i)$. $\eta=\zeta_8$ has been proven in \cite{sheng} not to be a norm, which guarantees that $\boldsymbol{\Xi}_k$ has a non zero determinant.

Let $\theta=\frac{1+\sqrt{5}}{2}$, $\sigma: \theta \mapsto \bar{\theta}=\frac{1-\sqrt{5}}{2}$ and the ring of integers of $\mathbb{K}$ $\mathcal{O}_{\mathbb{K}}=\{a+b\theta | a,b \in \mathbb{Z}[\zeta_8]\}$. Let $\alpha=1+i-i\theta$ and $\bar{\alpha}=1+i-i\bar{\theta}$. User's $k$ codeword $\boldsymbol{X}_k$ is
\begin{equation}
\boldsymbol{X}_k=\left[\begin{array}{cc}
\boldsymbol{\Xi}_k & \tau(\boldsymbol{\Xi}_k)\end{array}\right]\label{eq:codeword-k-2}
\end{equation}
where $\tau$ changes $\zeta_8$ into $-\zeta_8$ and $\boldsymbol{\Xi}_k$ defined in (\ref{eq:xi}) with $s_{kj}$ denoting the $j^{\mathrm{th}}$ QAM information symbol of user $i$. Finally, we get the equivalent codeword matrix of $\mathcal{C}_{(2,2,\boldsymbol{\Gamma})}$

\begin{equation}
\boldsymbol{X}=\left[\begin{array}{cc}
\boldsymbol{\Xi}_1 & \tau(\boldsymbol{\Xi}_1)\\
\boldsymbol{\Gamma}\boldsymbol{\Xi}_2 & \tau(\boldsymbol{\Xi}_2)\end{array}\right]\label{eq:codeword-eq}
\end{equation}
with 
\begin{equation}
\boldsymbol{\Gamma} = \left[\begin{array}{cc}
0 & 1 \\
i & 0
\end{array}\right].\label{eq:gamma}
\end{equation}
\begin{thm}
$\mathcal{C}_{(2,2,\boldsymbol{\Gamma})}$ achieves the outage DMT of the $K=2, n_t=2, n_r$ MIMO MAC channel. \end{thm}
\begin{proof} (sketch) If one of the users (say user $2$) is not in error, then the receiver can cancel the signal it receives from this user and the system is equivalent to a single-user $2\times n_r$ MIMO system. User $1$ transmits $\boldsymbol{X}_1$ given in (\ref{eq:codeword-k-2}) which is simply obtained by rotating $\boldsymbol{\Xi}_1$, hence, it is equivalent to the Golden Code which is known to be a DMT achievable space-time block code for $n_t=2$ transmit antennas and $n_r \geq 2$ receive antennas \cite{GC}.

If both users are in error, the system is equivalent to a $4 \times n_r$ MIMO channel and the transmitted codewords are given in (\ref{eq:codeword-eq}). Determinant of these codewords is
$$
\det\boldsymbol{X}=\det\left(\tau\left(\boldsymbol{\Xi}_2\right)-\boldsymbol{\Gamma}\boldsymbol{\Xi}_2\boldsymbol{\Xi}^{-1}_1\tau\left(\boldsymbol{\Xi}_1\right)\right)\det\boldsymbol{\Xi}_1
$$
Since $\tau\left(\boldsymbol{\Xi}_2\right)-\boldsymbol{\Gamma}\boldsymbol{\Xi}_2\boldsymbol{\Xi}^{-1}_1\tau\left(\boldsymbol{\Xi}_1\right)$ is in a division algebra, we get $\det\boldsymbol{X}=0$ \textit{iff}
$$
\tau\left(\boldsymbol{\Xi}_2\right)-\boldsymbol{\Gamma}\boldsymbol{\Xi}_2\boldsymbol{\Xi}^{-1}_1\tau\left(\boldsymbol{\Xi}_1\right)=0
$$
which gives
\begin{equation}
\boldsymbol{\Gamma} \Theta=\tau (\Theta)\label{Gamma-neq}
\end{equation}

for some $\Theta=\Xi_2 \Xi^{-1}_1 \in \mathcal{A}$. One solution that does not verify equation (\ref{Gamma-neq}) is to choose $\boldsymbol{\Gamma}$ to be a transcendantal scalar as we explained in the general case. But we can easily check that eq. (\ref{Gamma-neq}) is also not verified for $\boldsymbol{\Gamma}$ given in eq. (\ref{eq:gamma}). This condition is sufficient for our code to achieve the DMT \cite{Helmut-new}. 
\end{proof}

\section{Simulations}
In this section, we provide numerical results obtained by Monte-Carlo simulations. We assume that the power is allocated equally among all the users so that no a-priori advantage is given to any transmitter-receiver link over another one. We first present the outage performance of the considered channel. The performance of the proposed coding scheme is then measured by the word error rate (WER) \textit{vs} received $\mathsf{SNR}$ and compared to the time sharing scheme where the channel is shared among the users in an orthogonal multiple-access manner.

\subsection{Decoding Algorithm}
At the receiver side, we use a minimum mean-square error decision feedback equalizer MMSE-DFE preprocessing combined with lattice decoding as a way to tackle the problem of the rank deficiency resulting from $n_r$ being smaller than $K \times n_t$. In \cite{ElGamal-search}, it is shown that an appropriate combination of left, right preprocessing and lattice decoding, yields significant saving in complexity with very small degredation with respect to the ML performance. More precisely, left preprocessing modifies the channel matrix and the noise vector such that the resulting closest lattice point search has a much better conditioned channel matrix. Moreover, right preprocessing is used to change the lattice basis such that it becomes more convenient for the searching stage. 

\subsection{Numerical results}
\begin{figure}[ht]
\noindent \begin{centering}
\includegraphics[scale=0.49,angle=270]{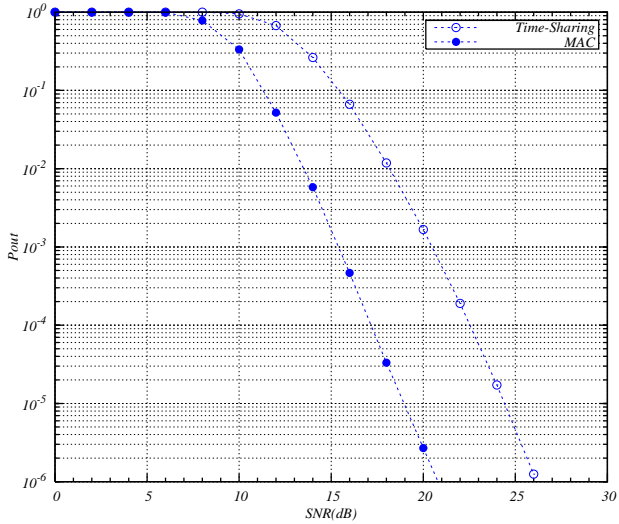}
\par\end{centering}
\caption{\label{fig:po-34}Outage performance of two-user MAC with 2 transmit antennas per user and three receive antennas, R=4 BPCU.}
\end{figure}

\begin{figure}[ht]
\noindent \begin{centering}
\includegraphics[scale=0.49,angle=270]{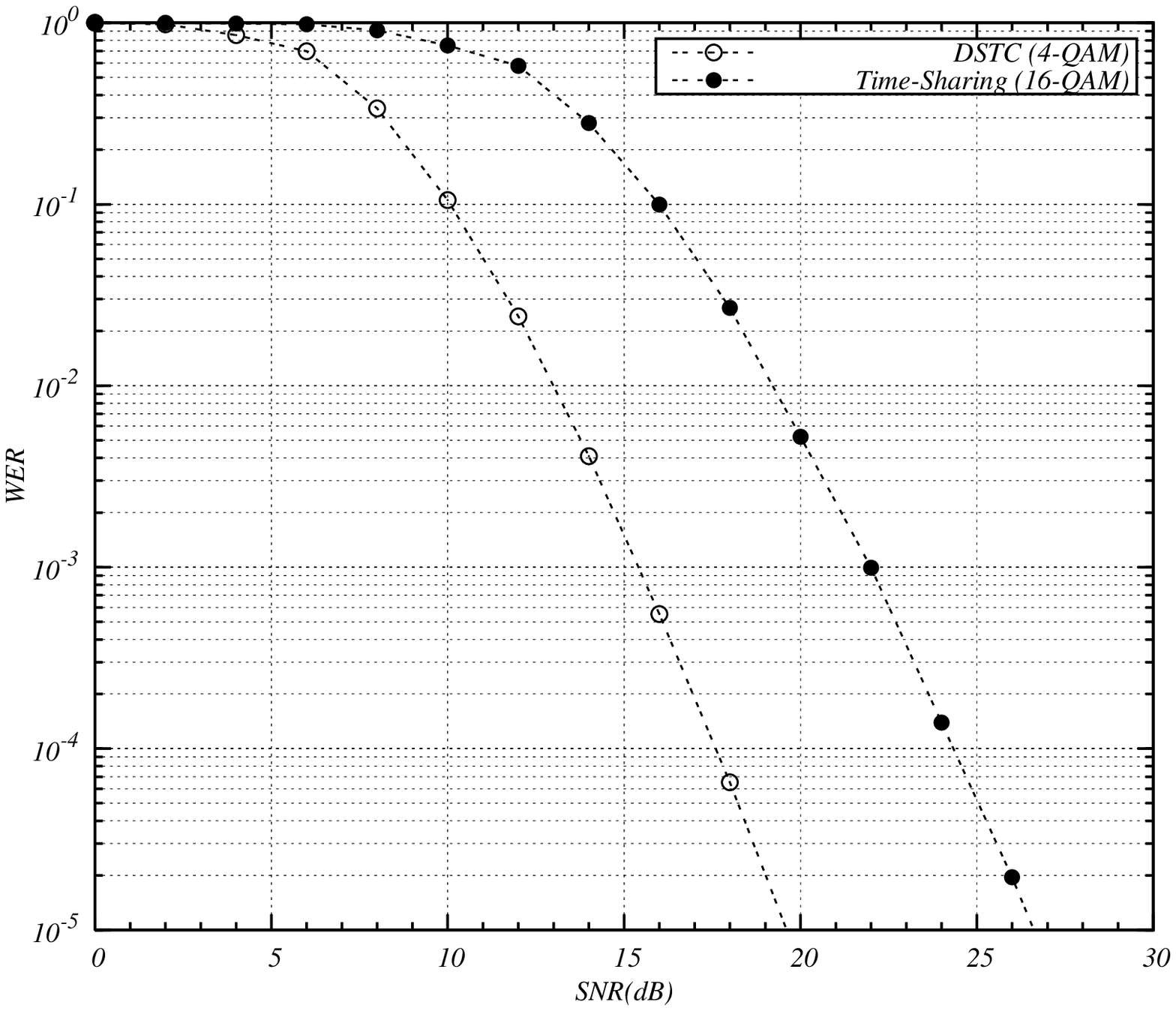}
\par\end{centering}
\caption{\label{fig:pe-34}Performance of the Space-Time Code designed for the two-user MAC with 2 transmit antennas per user, three receive antennas, 4-QAM.}
\end{figure}

\begin{figure}[ht]
\noindent \begin{centering}
\includegraphics[scale=0.49,angle=270]{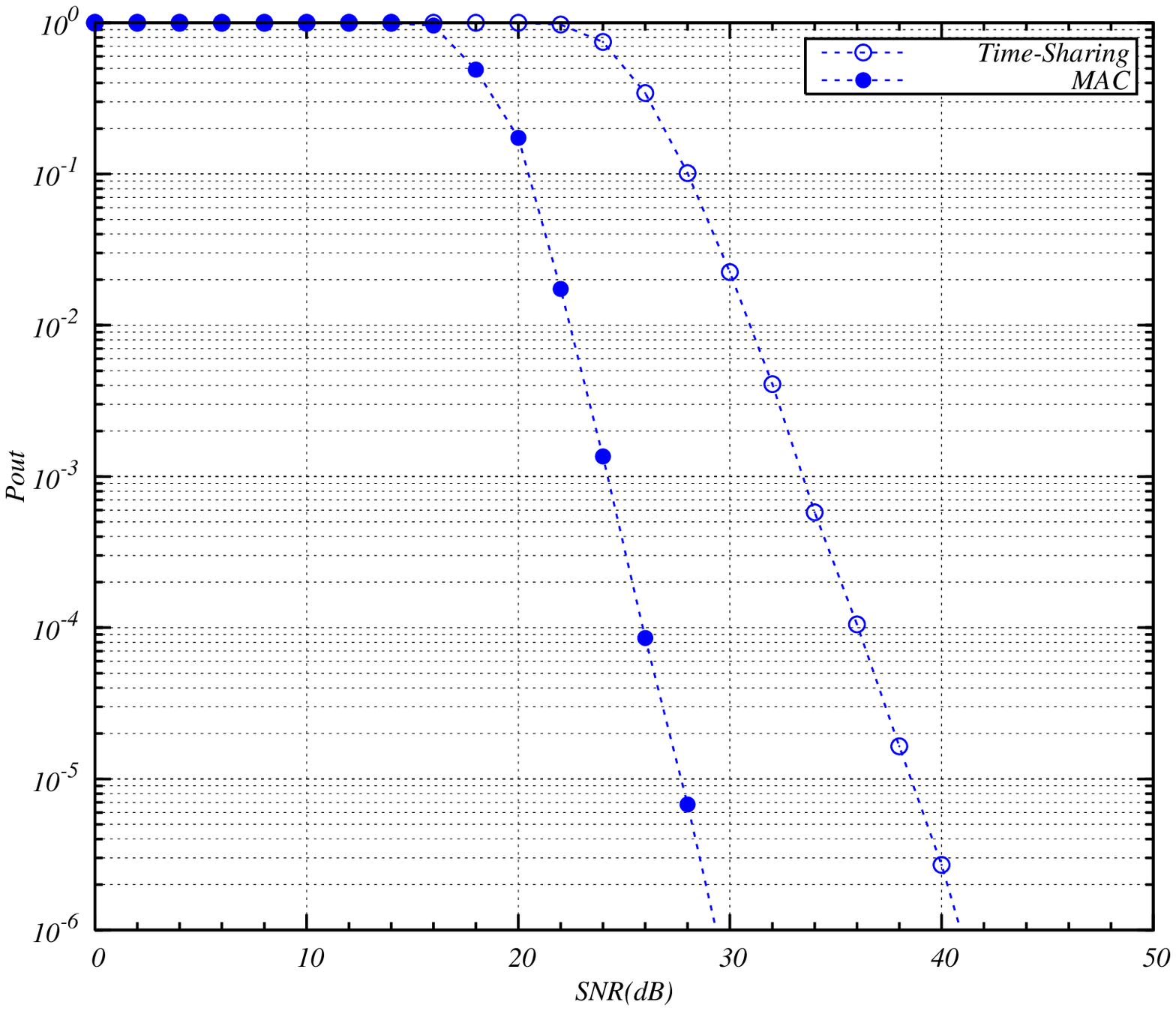}
\par\end{centering}
\caption{\label{fig:po-38} Outage performance of two-user MAC with 2 transmit antennas per user and three receive antennas, R=8 BPCU.}
\end{figure}

\begin{figure}[ht]
\noindent \begin{centering}
\includegraphics[scale=0.49,angle=270]{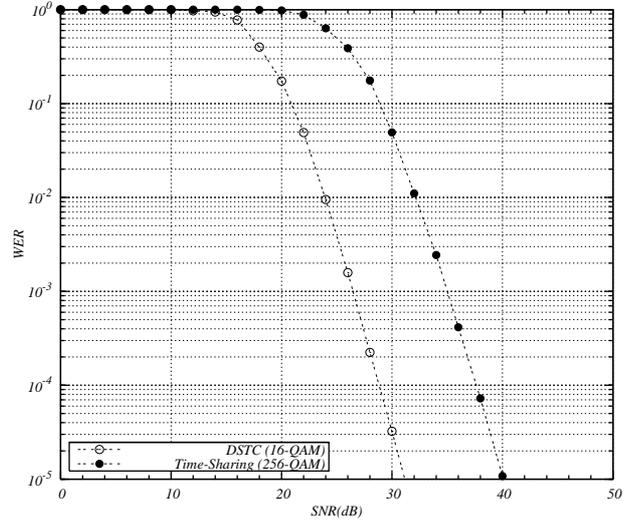}
\par\end{centering}
\caption{\label{fig:pe-316}Performance of the Space-Time Code designed for the two-user MAC with 2 transmit antennas per user, three receive antennas, 16-QAM.}
\end{figure}

We consider the two-user two-transmit antennas MAC with $n_r=3$ receive antennas. Outage performances for different spectral efficiencies are first illustrated in Figures \ref{fig:po-34} and \ref{fig:po-38} (4-BPCU and 8-BPCU, respectively). Coded schemes performances are shown in Figures \ref{fig:pe-34} and \ref{fig:pe-316}. Compared to the time sharing scheme, the proposed code achieves the same diversity order, 6, but offers a significant performance gain that depends on the spectral efficiency, $R$. In order to highlight this dependence on $R$, users information symbols are carved from different QAM constellations, \textit{e.g.} 4-QAM and a 16-QAM for the coded scheme (16-QAM and 256-QAM, repectively for the time-sharing scheme). At WER = $10^{-4}$, a gain of $6$ dB is observed when a 4-QAM constellation is considered. When we increase the spectral efficiency (16-QAM), this gain increases to $9$ dB. Interestingly, compared to the outage performance of the channel, the same behavior can be observed. This proves numerically the optimality of the proposed coding scheme.


\section{Conclusion}
In this paper, the multiantenna Multiple Access Channel with no Channel Side Information at the transmitters is considered. We propose a new construction of distributed space-time block codes that achieve the optimal DMT of the $K$-user MIMO-MAC. As an example, we present the special case of a two-user MAC with two transmit antennas per user. In order to overcome the rank deficiency, source of inefficiency of the well-known classical decoding approach, we used the MMSE-DFE preprocessing combined with the lattice decoding. Simulation results show that the new codes offer a significant performance gain compared to the time sharing scheme.



\begin{thebibliography}{1}
\bibitem{Zheng-1}
L.~Zheng and D.~Tse, ``Diversity and {M}ultiplexing: {A} fundamental tradeoff
  in multiple-antenna channels,'' {\em {IEEE} Trans. Inform. Theory}, vol.~49,
  pp.~1073--1096, May 2003.

\bibitem{DMTMAC}
D.~N. Tse, P.~Viswanath, and L.~Zheng, ``Diversity and multiplexing tradeoff in
  multiple-access channels,'' {\em IEEE Trans. Inform. Theory}, vol.~50,
  pp.~1859--1874, September 2004.

\bibitem{elgamal-mac}
Y.~Nam and H.~{El Gamal}, ``On the optimality of lattice coding and decoding in
  multiple access channels,'' in {\em Proceedings of ISIT 2007}, June 2007,
  Nice.

\bibitem{ElGamal-search}
A.D.~Murugan, H.~{El Gamal}, M.~O. Damen, and G.~Caire, ``A unified framework for tree search decoding: rediscovering the sequential decoder,'' {\em
  {IEEE} Trans. Inform. Theory}, vol.~52, pp.~933--953, March 2006.

\bibitem{ElGamal-4}
H.~{El Gamal}, G.~Caire, and M.~O. Damen, ``Lattice coding and decoding achieve
  the optimal diversity-vs-multiplexing tradeoff of {MIMO} channels,'' {\em
  {IEEE} Trans. Inform. Theory}, vol.~50, pp.~968--985, June 2004.

\bibitem{helmut}
M.~G\"artner and H.~B\"olcskei, ``Multiuser space-time/frequency code design,'' in
  {\em Proceedings of ISIT 2006, Seattle}, July 2006.

\bibitem{Helmut-new}
M.~G\"artner and H.~B\"olcskei, ``Multiuser space-time code design,'' Personnal Communication.

\bibitem{sheng}
S.~Yang and J.C.~Belfiore, ``Perfect space-time block codes for parallel MIMO channels,'' in
  {\em Proceedings of ISIT 2006, Seattle}, July 2006.

\bibitem{gallager}
R.~G.~Gallager, ``A perspective on multiaccess channels,'' in
  {\em{IEEE} Trans. Inform. Theory}, vol.~31, pp.~124--142, March 1985.

\bibitem{Nous_IZS}
M.~Badr and J.-C~Belfiore, ``Distributed space-time codes for the non cooperative Multiple-Access Channel,'' in {\em IEEE International Zurich Seminar on Communications}, ETH Zurich, Switzerland, March 2008.

\bibitem{GC}
J.-C. Belfiore, G.~Rekaya, and E.~Viterbo, ``The {Golden} code: A 2 $\times$ 2
  full-rate space-time code with non-vanishing determinants,'' {\em IEEE Trans.
  Inform. Theory}, vol.~51, pp.~1432--1436, April 2005.

\bibitem{booklet}
F.~Oggier, J.-C.~Belfiore and E.~Viterbo (2007), ``Cyclic Division Algebras: A Tool for Space-Time Coding,'' \textit{Foundations and Trends in Communications and Information Theory}, Vol. 4, No 1, pp 1-95, 2007. 

\bibitem{elia}
P.~Elia, B. A.~Sethuraman, and P.~Kumar, ``Perfect space-time codes with minimum and non-minimum delay for any number of antennas,'' {\em IEEE Trans. Inform. Theory}, vol.~1, pp.~722--727, June 2005.

\end{thebibliography}
\end{document}